\documentclass[aps,prd,twocolumn,showpacs,preprintnumbers,amsmath,amssymb,showpacs,showkeys]{revtex4}
\usepackage{graphicx}
\usepackage{dcolumn}
\usepackage{bm}
\usepackage{amssymb}
\usepackage{indentfirst}
\usepackage{epsfig}
\usepackage{epsf}
\usepackage{graphicx}
\usepackage{slashbox}

\newcommand{\eq}{\begin{eqnarray}}
\newcommand{\en}{\end{eqnarray}}

\begin{document}
       
\title{Considering anomalous dimensions in AdS/QCD models}

\author{Alfredo Vega$^{1}$\footnote{Talk given at Light Cone 2010: Relativistic Hadronic and Particle Physics, June 14-18, 2010, Valencia, Spain} and
        Ivan Schmidt$^{1}$
\vspace*{1.2\baselineskip}}

\affiliation{$^{1}$Departamento de F\'\i sica y Centro Cient\'ifico y 
Tecnol\'ogico de Valpara\'iso,\\  
Universidad T\'ecnica Federico Santa Mar\'\i a,\\
Casilla 110-V, Valpara\'\i so, Chile \\
\vspace*{.2\baselineskip} \\
\vspace*{.8\baselineskip}}

\date{\today}

\begin{abstract}

We discuss an AdS / QCD model that consider anomalous dimensions. The effect of this kind of dimensions is considered as a mass term that depend on holographical coordenate for duals modes in bulk.

\end{abstract}



\maketitle

\section{Introduction}

According common AdS / CFT dictionary, operators in a field theory has dimensions related to conformal dimension of dual modes that are propagating inside the bulk, but in general these dimensions are disregarded. A property of anomalous dimensions is his scale dependence, and this can be considered in AdS side using a mass term that depend on holographical coordinate for dual modes related to operator with anomalous dimensions. In this talk we discuss this alternative in AdS / QCD models that consider explicitly effect of chiral symmetry breaking in the lagrangian \cite{Erlich:2005qh, Da Rold:2005zs}

One kind of such models, known as hard wall, allows
to break chiral symmetry both spontaneously and explicitly in an
independent way, but the hadronic spectra calculated in this case
turns out to be not good. This situation can be improved by the
introduction a dilaton field, which in many articles is considered
quadratic in the holographical coordinate z \cite{Karch:2006pv}. The
obtained spectra has Regge behavior, but unfortunately it is not
possible now to break chiral symmetry explicitly and spontaneously
\cite{Colangelo:2008us}.

The  satisfactory implementation of chiral symmetry breaking,
without sacrificing the hadronic spectra, is a problem that has
attracted much interest lately. Examples of these kind of efforts
can be found in \cite{Zuo:2009dz, Gherghetta:2009ac, Kwee:2007nq,
Sui:2009xe, Zhang:2010tk}. In this talk we show a different alternative,
since we consider that the mass for modes propagating inside the
bulk can present a dependence on the holographical coordinate z \cite{Vega:2010ne},
which could be due to the fact that operators associated to these
modes might have an anomalous dimension \cite{Cherman:2008eh}.

It is possible to find references to z dependent masses in the
literature \cite{Cherman:2008eh}, where the authors suggest that the
anomalous dimension of operators can be translated into z dependent
masses for dual modes of these operators. This idea was used
successfully in \cite{Vega:2008te}, where an holographic model
without explicit chiral symmetry breaking was considered, and which
can reproduce the hadronic spectrum for spin 1/2 and 3/2 baryons
with with an arbitrary number of constituents. As is known for spin 1/2 case, a
dilaton field can not improve hard wall models, because this field
is factorized from the equation that gives us the spectra \cite{Kirsch:2006he, Vega:2008te}. Other
work related to mass varying in the bulk can be found in papers such
as \cite{Forkel:2007cm, Forkel:2007zz, dePaula:2009za,
deTeramond:2008ht}.

In this talk we consider an AdS / QCD model that takes into
account effects of chiral symmetry breaking, with z dependent scalar
mode masses. Thus we show that it is
possible to build a model that incorporates both spontaneous and
explicit chiral symmetry breaking, and which includes variable
masses. In the model presented, to certain set of parameter, we obtain that lighter scalar meson  has a mass lower that of the Pion, contradicting some properties of QCD \cite{Weingarten:1983uj, Witten:1983ut},  but this problem, fortunately no show for all the cases, so the model discussed here can be considered as a complementary alternative treatment for this
problem, different to the effort developed in
\cite{Gherghetta:2009ac, Kwee:2007nq, Sui:2009xe}, where the authors
try to improve soft wall models by deforming the dilaton and/or the
metrics.

The work consists of the following parts. Section II is a brief
description of the model, where we write down the equations that
describe the vev and the scalar, vector and axial vector mesons in
the AdS side. In III we obtain a variable mass for the scalar modes.
In section IV we discuss how to fix the parameters involved in this
model, in order to obtain in section V the spectra with the
parameters of the previous section. Finally, section VI is dedicated
to expose the conclusions of this work.

\section{Model}

\begin{table}
\begin{center}
\caption{Field content and dictionary of the model.}
\begin{tabular}{ c c c | c c c | c c c | c c c | c c c }
  \hline
  \hline
  & 4D : \textit{O}(x) & & & 5D : $\Phi(x,z)$ & & & p & & & $\Delta$ & & & $m_{5}^{2} R^{2}$ & \\
  \hline
  & $\overline{q}_{L} \gamma^{\mu} t^{a} q_{L}$ & & & $A_{L \mu}^{a}$ & & & 1 & & & 3 & & & 0 & \\
  & $\overline{q}_{R} \gamma^{\mu} t^{a} q_{R}$ & & & $A_{R \mu}^{a}$ & & & 1 & & & 3 & & & 0 & \\
  & $\overline{q}_{R}^{\alpha} q_{L}^{\beta}$ & & & $\frac{1}{z} X$ & & & 0 & & & $3 + \delta$ & & & $m_{5}^{2}(z) R^{2}$ & \\
  \hline
  \hline
\end{tabular}
\end{center}
\end{table}

First, in a brief way we like to remember that dimensions of operator in a field theory involved in the AdS / CFT correspondence are related with conformal dimension of dual modes in AdS side. In general, operator dimensions in a theory like QCD has anomalous dimensions that run with energy scale, and for other side, conformal dimensions for AdS modes depend on mass of this modes. If we consider other part of adS / CFT dictionary that say us the holographical coordenate z is related with energy, the relationship between dimension for operators in the field theory and conformal dimension for gravity modes tell us the mass for modes propagating inside the bulk must be z dependent. 

We consider the most usual version of soft wall AdS / QCD models,
with the notation used in \cite{Gherghetta:2009ac}, which takes into
account an 5d AdS background defined by
\begin{equation}
 \label{Metrica}
 d s^{2} = \frac{R^{2}}{z^{2}} (\eta_{\mu \nu} dx^{\mu} dx^{\nu} + dz^{2}),
\end{equation}

where R is the AdS radius, the Minkowsky metric is $\eta_{\mu \nu} =
diag (-1, +1, +1, +1)$ and z is a holographical coordinate defined
in $0 \leq z < \infty$. Here we consider a usual quadratic
dilaton
\begin{equation}
 \label{Dilaton}
 \phi (z) = \lambda^{2} z^{2}.
\end{equation}
To describe chiral symmetry breaking in the mesonic sector in the 5d
AdS side, the action considers $SU(2)_{L} \times SU(2)_{R}$  gauge fields
and a scalar field X. Such action is given by
\begin{equation}
 S_{5} = - \int d^{5}x \sqrt{-g} e^{-\phi(z)} Tr \biggl[|DX|^{2} + m_{X}^{2} (z) |X|^{2} \nonumber
\end{equation}
\begin{equation}
\label{Accion}
 + \frac{1}{4 g_{5}^{2}} (F_{L}^{2} + F_{R}^{2})\biggr].
\end{equation}
This action shows explicitly that the scalar modes masses are z
dependent, which is the feature that distinguishes this model
from other AdS / QCD models with chiral
symmetry breaking.

In Table I the fields included in model are shown and also his relationship with
modes propagating in the bulk,  according to AdS / CFT dictionary.
Notice that the operator $q\overline{q}$  has an
anomalous dimension ($\delta$), which in turn produces a mass that
depends on z for the modes dual to this operator, in agreement with
\cite{Cherman:2008eh, Vega:2008te}.

Starting from (\ref{Accion}), the equations that describe the vev and the scalar, vector and axial vector mesons are
\begin{widetext}
\begin{equation}
 \label{vev}
 - z^{2} \partial_{z}^{2} v(z) + z ( 3 + 2 \lambda^{2} z^{2} ) \partial_{z} v(z) + m_{X}^{2} (z) R^{2} v(z) = 0.
\end{equation}
\begin{equation}
 \label{escalar}
 - \partial_{z}^{2} S_{n} (z) + \biggl( \frac{3}{z} + 2 \lambda^{2} z \biggr) \partial_{z} S_{n} (z) + \frac{m_{X}^{2} (z) R^{2}}{z^{2}} S_{n} (z) = M_{S}^{2} S_{n}(z),
\end{equation}
\begin{equation}
 \label{vector}
 - \partial_{z}^{2} V_{n} (z) + \biggl( \frac{1}{z} + 2 \lambda^{2} z \biggr) \partial_{z} V_{n} (z) = M_{V}^{2} V_{n}(z),
\end{equation}
\begin{equation}
 \label{vector axial}
 - \partial_{z}^{2} A_{n} (z) + \biggl( \frac{1}{z} + 2 \lambda^{2} z \biggr) \partial_{z} A_{n} (z) + \frac{R^{2} g_{5}^{2} v^{2} (z)}{z^{2}} A_{n} (z) = M_{A}^{2} A_{n}(z).
\end{equation}
\end{widetext}

Before discussing the phenomenology of this model,
it is necessary to know the precise form of $m_{X}^{2} (z)$,  something
we will now consider.

\section{Obtaining an expression for $m_{X}^{2} (z)$}

\begin{figure*}[ht]
  \begin{tabular}{c c c}
    \includegraphics[width=1.9 in]{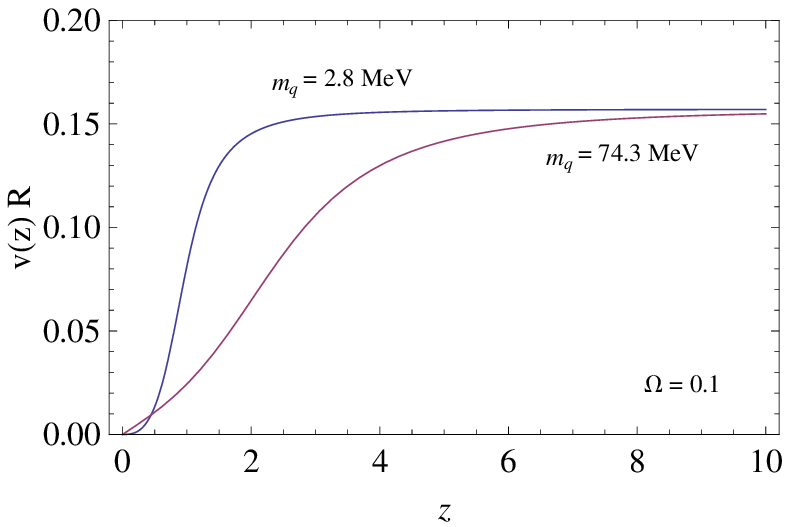}
    \includegraphics[width=1.9 in]{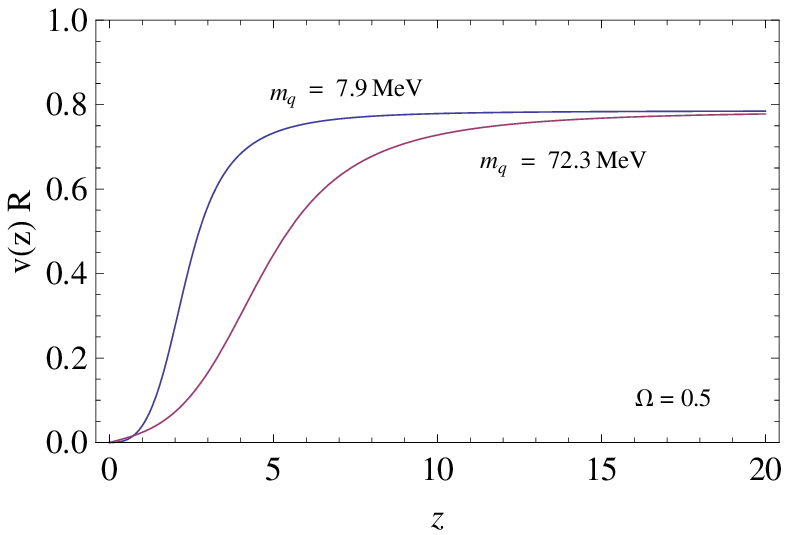}
    \includegraphics[width=1.9 in]{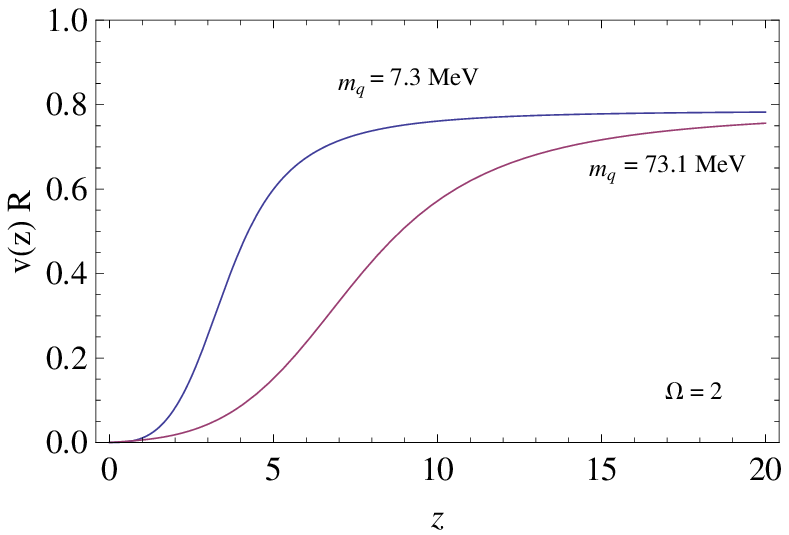} \\
    \includegraphics[width=1.9 in]{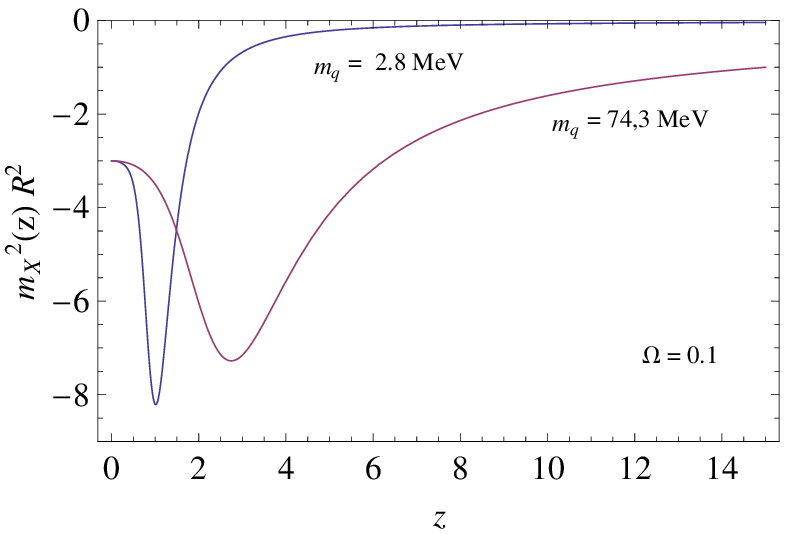}
    \includegraphics[width=1.9 in]{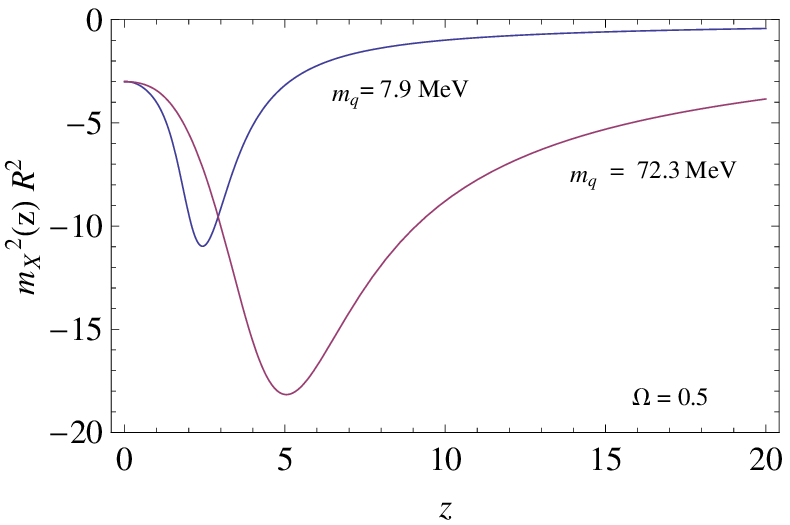}
    \includegraphics[width=1.9 in]{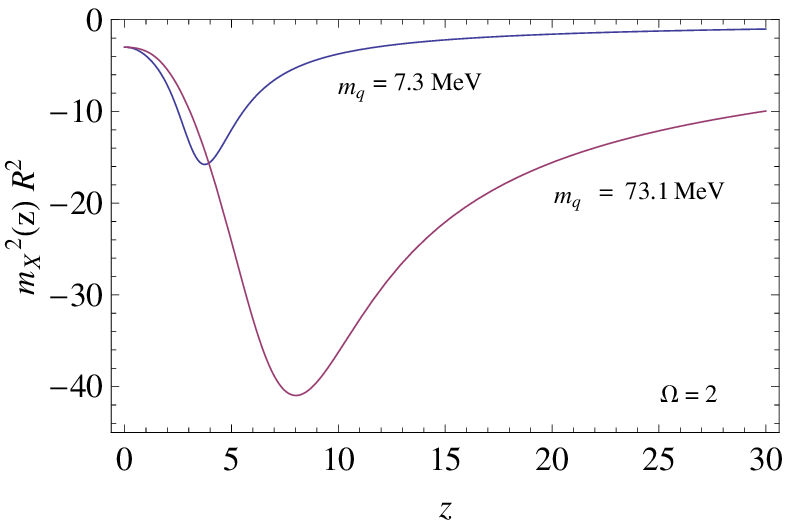}
  \end{tabular}
\caption{The upper graph shows  $v(z) R$, while in the lower graph the scalar modes masses as functions of z are given. All plots have been obtained with $\lambda = 0.4 GeV$, and $\Omega$ values used appear in each plot.}
\end{figure*}

The mass for scalar modes in the bulk,  $m_{X}^{2} (z)$, is obtained starting from (\ref{vev}),  although knowing first the function $v(z)$.  The behavior of this function is known in two limits.

First we consider the usual limit $z \rightarrow 0$, according to which
\begin{equation}
 \label{vev Cero}
v(z \rightarrow 0) = \alpha z + \beta z^{3},
\end{equation}
where the $\alpha$ and $\beta$ coefficient are associated with the quark mass and chiral condensate respectively.

The other limit in which we know the behavior of $v(z)$ is when $z
\rightarrow \infty$, and therefore we require that (\ref{vector
axial}) gives a Regge-like in this limit. In order to expose this
clearly, we change (\ref{vector axial}) using
\begin{equation}
 \label{TSchroAxial}
A_{n} (z) = \exp \biggl(\frac{1}{2} \int \biggl(\frac{1}{z} + 2 \lambda^{2} z\biggr) dz \biggr) a_{n} (z).
\end{equation}
This transformation converts our equation into a Schr\"odinger like equation, with a potential
\begin{equation}
 \label{PotencialAxial}
V_{A} (z) = \frac{3}{4 z^{2}} + \lambda^{4} z^{2} + \frac{R^{2} g_{5}^{2} v^{2} (z)}{z^{2}}.
\end{equation}

As is well known, soft wall models must reproduce spectra with Regge behavior when  $z \rightarrow \infty$, so the potential in this limit must look like
\begin{equation}
 \label{PotencialGeneral}
V (z) = a + b z^{2} + \frac{c}{z^{2}},
\end{equation}
and therefore $v(z \rightarrow \infty)$ can be: constant, linear or quadratic in z. In this work we only consider the first possibility, leaving for future work a more general discussion that considers all cases.

Knowing  the behavior of $v(z)$ both for $z \rightarrow 0$ as in $z \rightarrow \infty$, we choose an ansatz capable to reproduce both limits,
\begin{equation}
 \label{Ansatz vev}
v (z) = \frac{\Omega}{R} \arctan (A z + B z^{3}).
\end{equation}

Using this form for $v(z)$  in (\ref{vev}) allows us to get an expression for $m_{X}^{2} (z)$. Both $v(z)$ and $m_{X} (z)$ are shown in FIG 1 to different values for $\Omega$, where this is an arbitrary parameter and parameters A and B are related to quarks mass and chiral condensate and his values are fixed according to next section.

\section{Parameter setting}

The first parameter that we fix is $\lambda$, using data from the
spectrum. Specifically we consider a specific value for the Regge
slope, which in this kind of models with quadratic dilaton is $4
\lambda^{2}$. In \cite{Gherghetta:2009ac} a Regge slope is fixed
through radial excitations with $n \geq 3$, but in our case, since
the model has Regge behavior in the vector meson sector, we use a
value fixed by the lightest vector meson, so finally we choose
$\lambda = 0.400 GeV$ that allows us to obtain correct value masses
for vector mesons.

\begin{table*}[ht]
\begin{center}
\caption{Scalar meson spectra in MeV. Values for $\Omega$ are: (a) $\Omega = 0.1$, (b) $\Omega = 0.5$ y (c) $\Omega = 2$. As you can see in FIG 2, for each $\Omega$ there are two possible values for $m_{q}$. All masses are in MeV.}
\scalebox{0.9}[0.9]{
\begin{tabular}{ c | c | c | c | c | c | c | c | c }
  \hline
  \hline
  n & $f_{0} (Exp)$ & $f_{0} (a)$ & $f_{0} (a)$ & $f_{0} (b)$ & $f_{0} (b)$ & $f_{0} (c)$ & $f_{0} (c)$ & $f_{0} (Ref. [6])$ \\
    &               & $m_{q} = 2.8$ & $m_{q} = 74.3$ & $m_{q} = 7.9$ & $m_{q} = 72.3$ & $m_{q} = 7.3$ & $m_{q} = 73.1$ &    \\
  \hline
  0 & $550^{+250}_{-150}$ & 162 & 711 & 130 & 506 & 84 & 466 & 799 \\
  1 & $980 \pm 10$ & 1151 & 1179 & 1130 & 1036 & 1020 & 932 & 1184 \\
  2 & $1350 \pm 150$ & 1416 & 1444 & 1411 & 1357 & 1354 & 1253 & 1466 \\
  3 & $1505 \pm 6$ & 1635 & 1659 & 1632 & 1597 & 1595 & 1511 & 1699 \\
  4 & $1724 \pm 7$ & 1823 & 1846 & 1823 & 1799 & 1796 & 1729 & 1903 \\
  5 & $1992 \pm 16$ & 1999 & 2014 & 1995 & 1976 & 1974 & 1918 & 2087 \\
  6 & $2103 \pm 8$ & 2156 & 2169 & 2152 & 2137 & 2135 & 2089 & 2257 \\
  7 & $2314 \pm 25$ & 2303 & 2314 & 2299 & 2286 & 2284 & 2244 & 2414 \\
  \hline
  \hline
\end{tabular}}
\end{center}
\end{table*}

The remaining parameters can be fixed using (\ref{Ansatz vev}), the expression chosen to describe the vev, which has the limits
\begin{equation}
 \label{Ansatz vev UV}
v (z \rightarrow 0) = \frac{\Omega}{R} A z + \frac{\Omega}{R} \biggl(-\frac{A^{3}}{3} + B \biggr) z^{3} + O(z^{5}),
\end{equation}
\begin{equation}
 \label{Ansatz vev IR}
v (z \rightarrow \infty) = \frac{\Omega \pi}{2 R} + O(z^{-3}).
\end{equation}

Comparing (\ref{Ansatz vev UV}) with the value established in the AdS / CFT dictionary, with the notation used in \cite{Gherghetta:2009ac}
\begin{equation}
 \label{vev UV}
v (z \rightarrow 0) = \frac{ m_{q} \zeta}{R} z + \frac{\sigma}{R \zeta} z^{3},
\end{equation}
The parameter $\zeta$  was introduced in \cite{Cherman:2008eh} to get the right normalization, and his value is   $\zeta = \sqrt{3}/(2 \pi)$. With this, the A and B parameters are given by
\begin{equation}
 \label{Parametro A}
A = \frac{\sqrt{3}}{2 \pi \Omega} m_{q}
\end{equation}
and
\begin{equation}
 \label{Parametro B}
B = \frac{2 \pi}{\sqrt{3} \Omega} \sigma + \frac{3 \sqrt{3}}{8 \pi^{3} \Omega^{3}} m^{3}_{q},
\end{equation}
where $m_{q}$ is the quark mass and $\sigma$ is the chiral condensate.

In order to finish the model description, it is necessary to specify the values for $m_{q}$ and $\sigma$, which are related by the GOR $m^{2}_{\pi} f^{2}_{\pi} = 2 m_{q} \sigma$, and therefore we need to fix only one of them. In this case we use $m_{\pi} = 140 MeV$ and $f_{\pi} = 92 MeV$, and we fix the quark mass using
\begin{equation}
 \label{Constante Decaimiento}
f^{2}_{\pi} = - \frac{1}{g^{2}_{5}} \lim_{\epsilon \rightarrow 0} \frac{\partial_{z} A_{0} (0,z)}{z} |_{z = \epsilon},
\end{equation}
where $A_{0} (0,z)$ is solution of (\ref{vector axial}), with $M^{2}_{A} = 0$, and the boundary conditions used are $A_{0}(0,0)=1$ and $\partial_{z} A_{0} (0,z \rightarrow \infty) = 0$.

As you can see in (\ref{vector axial}), $A_{0} (0,z)$ equation as a term that depend on $m_{q}$, so using (\ref{Constante Decaimiento}) we get $f_{\pi} (m_{q})$. This appear in FIG 2, and this show us for each $\Omega$ value, we have two possible quark masses. For $\Omega = 0.1$ we found $m_{q} = 2.8 MeV$ and $m_{q} = 74.3 MeV$; for $\Omega = 0.5$ we get $m_{q} = 7.9 MeV$ and $m_{q} = 72.3 MeV$ and when we use $\Omega = 2$ is obtained $m_{q} = 7.3 MeV$ and $m_{q} = 73.1 MeV$.

\begin{figure}[h]
\begin{center}
    \includegraphics[width=2.5 in]{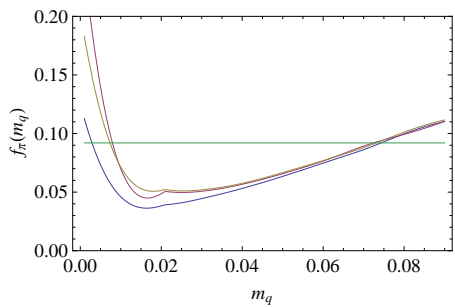}
\end{center}
\caption{The plot show the pion decay constant as $m_{q}$ function according to (\ref{Constante Decaimiento}). The horizontal line correspond to $f_{\pi} = 92 MeV$.}
\end{figure}

\section{Mesonic spectrum}

Having fixed the parameters of the model, we can calculate masses for some mesons, which correspond to eigenvalues in the equations (\ref{escalar}), (\ref{vector}) and (\ref{vector axial}). In this set of equations, only (\ref{vector}) can be solved analytically. For this reason we prefer to change all equations into Schr\"odinger like ones, and later solve numerically (\ref{escalar}) and (\ref{vector axial}) using a MATHEMATICA code called schroedinger.nb \cite{Lucha:1998xc},  which was adapted to our potentials.

\begin{table}[ht]
\begin{center}
\caption{Vector mesons spectra in MeV.}
\begin{tabular}{ c c c | c c c | c c c | c c c }
  \hline
  \hline
  & n & & & $\rho (Exp)$ & & & $\rho (Model)$ & & & $\rho (Ref.[6])$ & \\
  \hline
  & 0 & & & $775.5 \pm 1$ & & & 800 & & & 475 & \\
  & 1 & & & $1282 \pm 37$ & & & 1131 & & & 1129 & \\
  & 2 & & & $1465 \pm 25$ & & & 1386 & & & 1529 & \\
  & 3 & & & $1720 \pm 20$ & & & 1600 & & & 1674 & \\
  & 4 & & & $1909 \pm 30$ & & & 1789 & & & 1884 & \\
  & 5 & & & $2149 \pm 17$ & & & 1960 & & & 2072 & \\
  & 6 & & & $2265 \pm 40$ & & & 2117 & & & 2243 & \\
  \hline
  \hline
\end{tabular}
\end{center}
\end{table}

\begin{table*}[ht]
\begin{center}
\caption{Axial vector mesons spectra in MeV. Values for $\Omega$ are: (a) $\Omega = 0.1$, (b) $\Omega = 0.5$ and (c) $\Omega = 2$. The values for $m_{q}$ in the table. All masses are in MeV.}
\scalebox{0.9}[0.9]{
\begin{tabular}{ c | c | c | c | c | c | c | c | c }
  \hline
  \hline
   n & $a_{1} (Exp)$ & $a_{1} (a)$ & $a_{1} (a)$ & $a_{1} (b)$ & $a_{1} (b)$ & $a_{1} (c)$ & $a_{1} (c)$ & $a_{1} (Ref.[6])$ \\
     &               & $m_{q} = 2.8$ & $m_{q} = 74.3$ & $m_{q} = 7.9$ & $m_{q} = 72.3$ & $m_{q} = 7.3$ & $m_{q} = 73.1$ &    \\
  \hline
   0 & $1230 \pm 40$ & 864 & 825 & 1262 & 888 & 1495 & 909 & 1185  \\
   1 & $1647 \pm 22$ & 1170 & 1144 & 1388 & 1231 & 2331 & 1325 & 1591 \\
   2 & $1930^{+39}_{-70}$ & 1412 & 1394 & 1579 & 1468 & 2544 & 1649 & 1900 \\
   3 & $2096 \pm 122$ & 1619 & 1606 & 1755 & 1668 & 2660 & 1910 & 2101 \\
   4 & $2270^{+55}_{-40}$ & 1803 & 1794 & 1917 & 1846 & 2770 & 2117 & 2279 \\
  \hline
  \hline
\end{tabular}}
\end{center}
\end{table*}

\section{Conclusions}

The possibility of incorporating chiral symmetry breaking in soft
wall models, introducing a dependence on the holographical
coordinate in the mass for models propagating inside the bulk, was
studied. This idea could be considered as a complement to other
mechanisms that try to solve this problem introducing changes in the
dilaton field, changes in the metric, or introducing a cubic or
quartic term for scalars in the action \cite{Gherghetta:2009ac,
Kwee:2007nq, Sui:2009xe, Zhang:2010tk}.

The model considered here uses a usual quadratic dilaton and a AdS metric, and considers an expression   for  $v(z)$ that it is able to reproduce the expected behavior in the UV and IR limits. For certain choice of parameters is obtained that the mass of the lightest scalar meson is less than the mass of the Pion, contradicting a theorem well-established QCD, which fortunately for the model is not so in all cases, then it is possible to obtain mesonic masses that in general are in good agreement with data.

In light of the results presented in this paper, we think that the
introduction of a mass that varies with z inside the bulk can be
considered as a complementary alternative in order to build AdS /
QCD models that take into account effects of chiral symmetry
breaking. In some cases the spectra that we get is poor, so a z
dependent mass like we are presenting here clearly cannot solve all
problems, but anyway we think is an interesting complementary
alternative, because z dependent masses can be associated to dual
modes of operators with anomalous dimensions, allowing the
introduction into this kind of models of an important  QCD quantity,
which is usually not considered when people build AdS / QCD models.

\begin{acknowledgments}

Work supported by Fondecyt (Chile) under Grants No. 3100028 and 1100287.

\end{acknowledgments}


\begin{thebibliography}{99}

\bibitem{Erlich:2005qh}
  J.~Erlich, E.~Katz, D.~T.~Son and M.~A.~Stephanov,
  Phys.\ Rev.\ Lett.\  {\bf 95}, 261602 (2005)
  [arXiv:hep-ph/0501128].

\bibitem{Da Rold:2005zs}
  L.~Da Rold and A.~Pomarol,
  Nucl.\ Phys.\  B {\bf 721}, 79 (2005)
  [arXiv:hep-ph/0501218].

\bibitem{Karch:2006pv}
  A.~Karch, E.~Katz, D.~T.~Son and M.~A.~Stephanov,
  Phys.\ Rev.\  D {\bf 74}, 015005 (2006)
  [arXiv:hep-ph/0602229].

\bibitem{Colangelo:2008us}
  P.~Colangelo, F.~De Fazio, F.~Giannuzzi, F.~Jugeau and S.~Nicotri,
  Phys.\ Rev.\  D {\bf 78}, 055009 (2008)
  [arXiv:0807.1054 [hep-ph]].

\bibitem{Zuo:2009dz}
  F.~Zuo,
  arXiv:0909.4240 [hep-ph].

\bibitem{Gherghetta:2009ac}
  T.~Gherghetta, J.~I.~Kapusta and T.~M.~Kelley,
  Phys.\ Rev.\  D {\bf 79}, 076003 (2009)
  [arXiv:0902.1998 [hep-ph]].

\bibitem{Kwee:2007nq}
  H.~J.~Kwee and R.~F.~Lebed,
  Phys.\ Rev.\  D {\bf 77}, 115007 (2008)
  [arXiv:0712.1811 [hep-ph]].

\bibitem{Sui:2009xe}
  Y.~Q.~Sui, Y.~L.~Wu, Z.~F.~Xie and Y.~B.~Yang,
  Phys.\ Rev.\  D {\bf 81}, 014024 (2010)
  [arXiv:0909.3887 [hep-ph]].

\bibitem{Zhang:2010tk}
  P.~Zhang,
  arXiv:1003.0558 [hep-ph].
  
\bibitem{Vega:2010ne}
  A.~Vega and I.~Schmidt,
  arXiv:1005.3000 [hep-ph].

\bibitem{Cherman:2008eh}
  A.~Cherman, T.~D.~Cohen and E.~S.~Werbos,
  Phys.\ Rev.\  C {\bf 79}, 045203 (2009)
  [arXiv:0804.1096 [hep-ph]].

\bibitem{Vega:2008te}
  A.~Vega and I.~Schmidt,
  Phys.\ Rev.\  D {\bf 79}, 055003 (2009)
  [arXiv:0811.4638 [hep-ph]].
  
\bibitem{Kirsch:2006he}
  I.~Kirsch,
  JHEP {\bf 0609}, 052 (2006)
  [arXiv:hep-th/0607205].

\bibitem{Forkel:2007cm}
  H.~Forkel, M.~Beyer and T.~Frederico,
  JHEP {\bf 0707}, 077 (2007)
  [arXiv:0705.1857 [hep-ph]].

\bibitem{Forkel:2007zz}
  H.~Forkel, M.~Beyer and T.~Frederico,
  Int.\ J.\ Mod.\ Phys.\  E {\bf 16}, 2794 (2007).

\bibitem{dePaula:2009za}
  W.~de Paula and T.~Frederico,
  arXiv:0908.4282 [hep-ph].

\bibitem{deTeramond:2008ht}
  G.~F.~de Teramond and S.~J.~Brodsky,
  Phys.\ Rev.\ Lett.\  {\bf 102}, 081601 (2009)
  [arXiv:0809.4899 [hep-ph]].
  
\bibitem{Weingarten:1983uj}
  D.~Weingarten,
  Phys.\ Rev.\ Lett.\  {\bf 51}, 1830 (1983).

\bibitem{Witten:1983ut}
  E.~Witten,
  Phys.\ Rev.\ Lett.\  {\bf 51}, 2351 (1983).

\bibitem{Lucha:1998xc}
  W.~Lucha and F.~F.~Schoberl,
  Int.\ J.\ Mod.\ Phys.\  C {\bf 10}, 607 (1999)
  [arXiv:hep-ph/9811453].


\end{thebibliography}
\end{document}